\documentclass[journal]{IEEEtran}
\usepackage{epsfig}
\usepackage{graphicx}
\usepackage{amsmath}
\usepackage{amssymb}
\usepackage{amsmath} 
\usepackage{amsmath} 
\usepackage{algorithmic}
\usepackage{algorithm}
\usepackage{subfigure}
\usepackage{multirow}
\usepackage{url}
\usepackage{color}
\usepackage{hyperref}
\usepackage{balance}
\usepackage{setspace}
\usepackage{booktabs}
\usepackage{threeparttable}
\usepackage{float}

\newcommand{\etal}{{\em et al.}}       
\newcommand{\eg}{{\em e.g.}}           
\newcommand{\ie}{{\em i.e.}}           
\newcommand{\etc}{{\em etc}}         
\ifCLASSOPTIONcompsoc
  \usepackage[nocompress]{cite}
\else
  \usepackage{cite}
\fi

\ifCLASSINFOpdf
\else
\fi

\begin{document}
%
\title{Deep Symmetric Adaptation Network for Cross-modality Medical Image Segmentation}

%
%
%
%

\author{Xiaoting~Han$^\dagger$, Lei~Qi$^\dagger$, Qian~Yu, Ziqi~Zhou, Yefeng~Zheng, Yinghuan~Shi$^\star$, Yang~Gao 
\IEEEcompsocitemizethanks{
\IEEEcompsocthanksitem \emph{Xiaoting Han and Lei Qi are the co-first authors. Corresponding Author: Yinghuan Shi}.
\IEEEcompsocthanksitem Xiaoting Han, Ziqi Zhou, Yinghuan Shi, and Yang Gao are with the State Key Laboratory for Novel Software Technology, the National Institute for Healthcare Data Science, and the Collaborative Innovation Center of Novel Software Technology and Industrialization, Nanjing University, China. (e-mail: hxt@smail.nju.edu.cn, zhouzq@smail.nju.edu.cn, syh@nju.edu.cn, gaoy@nju.edu.cn)
\IEEEcompsocthanksitem Lei Qi is with the School of Computer Science and Engineering, and the Key Lab of Computer Network and Information Integration
(Ministry of Education), Southeast University, China. (e-mail: qilei@seu.edu.cn)
\IEEEcompsocthanksitem Qian Yu is with the School of Data and Computer Science, Shandong Women's University, China. (e-mail: yuqian@sdwu.edu.cn)
\IEEEcompsocthanksitem Yefeng~Zheng is with the Tencent Jarvis Lab, China. (e-mail: yefengzheng@tencent.com)
\IEEEcompsocthanksitem This work was supported by the National Key research and Development Plan (2019YFC0118300) and NSFC (61673203).
}
}




\IEEEtitleabstractindextext{%
\begin{abstract}
  Unsupervised domain adaptation (UDA) methods have shown their promising performance in the cross-modality medical image segmentation tasks. These typical methods usually utilize a translation network to transform images from the source domain to target domain or train the pixel-level classifier merely using translated source images and original target images.
  However, when there exists a large domain shift between source and target domains, we argue that this asymmetric structure could not fully eliminate the domain gap. 
  In this paper, we present a novel deep symmetric architecture of UDA for medical image segmentation, which consists of a segmentation sub-network, and two symmetric source and target domain translation sub-networks. 
  To be specific, based on two translation sub-networks, we introduce a bidirectional alignment scheme via a shared encoder and private decoders to simultaneously align features 1) from source to target domain and 2) from target to source domain, which helps effectively mitigate the discrepancy between domains.
  Furthermore, for the segmentation sub-network, we train a pixel-level classifier using not only original target images and translated source images, but also original source images and translated target images, which helps sufficiently leverage the semantic information from the images with different styles. 
  Extensive experiments demonstrate that our method has remarkable advantages compared to the state-of-the-art methods in both cross-modality Cardiac and BraTS segmentation tasks. 
\end{abstract}

}

\maketitle
\IEEEdisplaynontitleabstractindextext
\IEEEpeerreviewmaketitle
\bigskip
\bigskip

\begin{figure}
  \centering
  \subfigure[Examples of CT and MR images.]{
  \includegraphics[width=4cm]{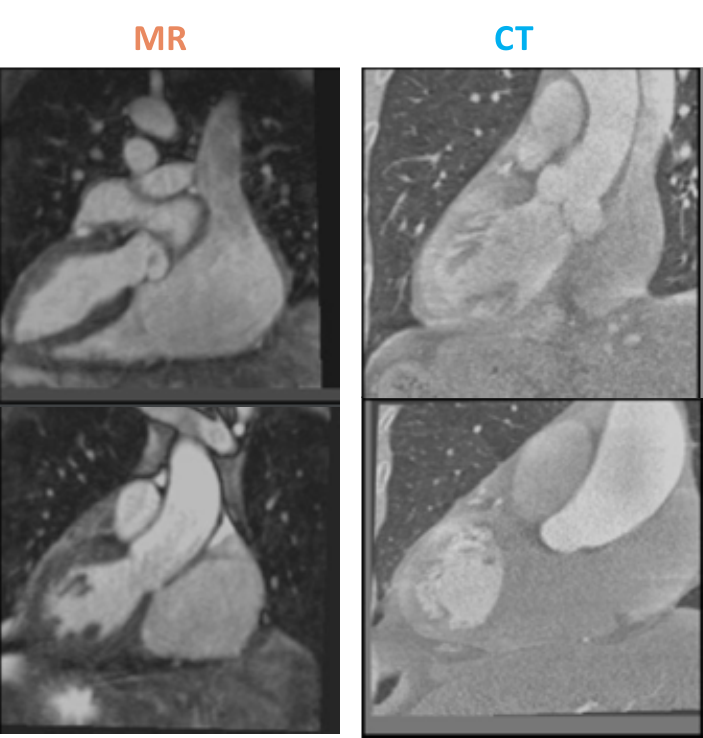}
  }
  \subfigure[Performance of different settings.]{
  \includegraphics[width=4cm,height=4cm]{figs/fig1-2.pdf}
  }
  \caption{Comparison of the models trained respectively on CT and MR images while evaluating on CT images. Note that, the smaller for ``ASD'' is better, and reverse on ``Dice''.}
  \label{fig1}
  \end{figure}

\IEEEraisesectionheading{\section{Introduction}}
\label{sec:introduction}
Deep convolutional neural networks have made significant progress in medical image segmentation task. In the task, there is a common assumption that training and test images are drawn from the same data distribution. However, in numerous real-world applications, especially in healthcare field, due to different acquisition parameters or various imaging modalities, a large gap of data distributions between training and test sets possibly occurs. This distribution gap usually causes a drastic performance drop during the deployment of trained model. For example, taking magnetic resonance imaging (MRI) and computed tomography (CT) in Fig.~\ref{fig1} as examples, we could observe there exists a large appearance variation between these two modalities. To quantitatively evaluate the performance drop caused by domain gap, we first train two respective segmentation models purely on CT or MR images, and then utilize these two models to segment a common CT image set, which denotes as ``CT to CT'' and ``MRI to CT'', respectively. It is obvious that a drastic performance degeneration happens in ``MRI to CT'' compared to ``CT to CT'', in terms of Dice and average surface distance (ASD). In this aspect, the model trained purely on MR images cannot directly generalize well on CT images due to this aforementioned domain gap.

  To reduce the performance drop across different modalities, a trivial yet straightforward way is to fine-tune the model pre-trained on source images with labeled target images. However, the labeling process which requires pixel-wise manual delineation on target images is often labor-intensive to achieve, especially when we have a massive number of images for labeling. To tackle this issue, unsupervised domain adaptation (UDA)-based segmentation methods~\cite{saito2018maximum, tsai2018learning, gatys2016image, vu2019dada} have aroused considerable attention in recent years. In UDA segmentation, given a fully labeled source domain, our goal is to segment another target domain without labels in a cross-domain manner.
  Thanks to the recent development of UDA-based segmentation, the performance of cross-domain segmentation has now been greatly improved.
  
  For these existing UDA-based segmentation methods, most of them focus on minimizing the differences between distributions of source and target domains from the \textit{image-translation} or \textit{feature-alignment} perspectives~\cite{chen2018road}. Specifically, for the image-translation perspective, a common way~\cite{zhu2017unpaired, choi2018stargan,chen2020unsupervised,chen2019synergistic} is to translate source domain images to the style of target domain images by image translation networks~\cite{bousmalis2017unsupervised, Shrivastava2017LearningFS}. The translated source images with their inherited ground truth labels from the source domain are utilized to train a target domain-oriented segmentation model in a supervised manner. However, due to the instability of generative adversarial network (GAN)-based methods~\cite{goodfellow2014generative, isola2017image, vu2019dada}, some inferior translations could destruct semantic information of original images, thus it may affect the segmentation accuracy. 
  
  For the feature-alignment perspective, several works~\cite{solomon2015convolutional, saleh2018effective, tzeng2015simultaneous} align distributions across domains in feature space to mitigate domain gap. 
  Recent studies~\cite{zhu2018penalizing, sankaranarayanan2018learning} propose to project the feature space to other compact spaces, such as image space, since latent space of semantic segmentation might be over-complex and high-dimensional, which should simultaneously encode appearance, shape and context, \etc. In this paper, we also apply adversarial learning in generated image space to mitigate distribution discrepancy. 
  
  Currently, the feature alignment is usually achieved by an independent network~\cite{zouunsupervised, chen2020unsupervised}. Instead, we choose to bidirectionally align feature distributions across domains using two symmetric translation sub-networks, which achieves a significant performance improvement as compared to applying an independent network.
  
  In this paper, we put forward a novel deep symmetric adaptation network including a segmentation sub-network, and source/target domain translation sub-networks. In this network, we utilize two translation sub-networks via sharing the encoder and using private decoders to realize the bidirectional alignment (\ie, from source to target and vice versa) of feature distributions between source and target domains, which can effectively alleviate the large discrepancy between domains. Moreover, all images from the translation sub-networks and raw source and target domains are used to train the segmentation sub-network, thus it helps sufficiently explore rich yet easily ignored semantic information. 
  We evaluate our method on cross-modality Cardiac~\cite{zhuang2016multi} and BraTS~\cite{menze2014multimodal} segmentation tasks.
  Extensive experiments reveal the superiority of our method, compared to current state-of-the-art (SOTA) methods. Moreover, ablation study highlights the effectiveness of each module developed in our method.
  Main contributions of this paper are summarized as:
  \begin{itemize}
      \item We develop a novel symmetric adaptation network for cross-modality medical image segmentation, which consists of a segmentation sub-network, source and target domain translation sub-networks.
      \item We propose a bidirectional alignment scheme over source and target translation sub-networks. Besides, all images of different styles from source and target domains are utilized to train the segmentation module.
      \item We conduct extensive experiments in the cardiac segmentation task: ``MRI to CT'' and ``CT to MRI''. We achieve a new state-of-the-art of 78.50\% and 66.45\% in mean Dice. We additionally evaluate our methods on the BraTS dataset, where our method outperforms other SOTA methods as well, achieving Dice of 67.18\%.
  \end{itemize}

  \section{Related Work}
  \label{sec:related-work}
  The existing UDA-based segmentation methods can be roughly divided into three categories according to different alignment perspectives, \ie, 1) feature-alignment-based methods, 2) image-translation-based methods and 3) joint learning methods of feature alignment and image translation.
  
  \subsection{Feature-alignment-based Methods}
  Many UDA methods pay attention to align distributions in the feature space, by reducing the distance metric between features extracted from source and target domains. Among them, maximum mean discrepancy (MMD) loss~\cite{gretton2012kernel, long2015learning, bousmalis2016domain}
  is a popularly-used distance metric. As its extension, some statistics of feature distributions such as covariance~\cite{sun2016deep, mancini2018boosting} are also utilized for alignment. Another prominent approach towards domain adaptation for image segmentation is adversarial learning~\cite{goodfellow2014generative, ganin2016domain}. Several works use adversarial learning for latent features between the encoder and decoder, which resorts to learn domain-invariant representations~\cite{chen2017no,luo2019taking,hong2018conditional}.
  
  Recent methods in literature~\cite{sankaranarayanan2018learning, zhu2018penalizing} perform adversarial learning in low-level space, \eg, image space, to achieve a compact embedding, since latent space of image segmentation might be over-complex and high-dimensional, including various visual cues, such as appearance and context.
  
  \subsection{Image-translation-based Methods}
  Inspired by GAN-based techniques~\cite{goodfellow2014generative,zhu2017unpaired}, some image-translation-based methods~\cite{bousmalis2017unsupervised, royer2020xgan, choi2018stargan} are developed to convert source images into target style-like images. Afterwards, these generated images inheriting ground truth of source images can be used for supervised training of target segmentation network.
  For example, the PixelDA method~\cite{bousmalis2017unsupervised} addresses domain adaptation by translating source images to target domain, thereby obtaining a simulated training set for target images.
  Huo \etal~\cite{huo2018synseg} first adopt a synthesis network to generate target-like images. Then they feed these generated images to a segmentation network to perform supervised training with source domain labels.
  However, due to instability of GAN-based methods, semantic information might be destructed during generating cross-domain images.

  \subsection{Joint Learning Methods}
  Recently, some hybrid works integrate image translation and feature alignment to better mitigate the domain shift~\cite{hoffman2018cycada,zhang2018fully,li2019bidirectional}. For example,
  Chang \etal~\cite{chang2019all} propose a domain invariant structure extraction (DISE) framework. This framework disentangles images into domain-invariant structure and domain-specific texture representation. It further translates image across domains and enables label transfer to improve the segmentation performance.
  Appearance adaptation networks (AAN) and representation adaptation networks (RAN) are combined in~\cite{zhang2018fully}. The former adapts source-domain images to the ``style'' of the target domain, and the latter attempts to learn domain-invariant representations.
  In medical image segmentation, Chen \etal~\cite{chen2019synergistic,chen2020unsupervised} present a novel unsupervised domain adaptation framework, namely synergistic image and feature alignment (SIFA). This method considers the case of severe domain shift in cross-modality medical images, and synergistically merges the feature alignment and image translation into a unified network.
  
  However, there are two disadvantages in the aforementioned methods. One is that aligning feature distributions via an independent network is difficult, especially for the case when distribution discrepancy is large. Another is that semantic information of original images might be destructed during generating cross-domain images.
  In response to these two shortcomings, we focus on the feature alignment using two symmetric translation sub-networks to make up for semantic knowledge for image-translation with source images. Furthermore, we explore more common semantic information using adversarial losses between not only original source and target images but also translated source and target images in the semantic space.

  \section{Method}
  \label{sec:method}
  \begin{figure*}
    \centering
    \setlength{\belowcaptionskip}{-1cm}
    \includegraphics[width=0.88\textwidth]{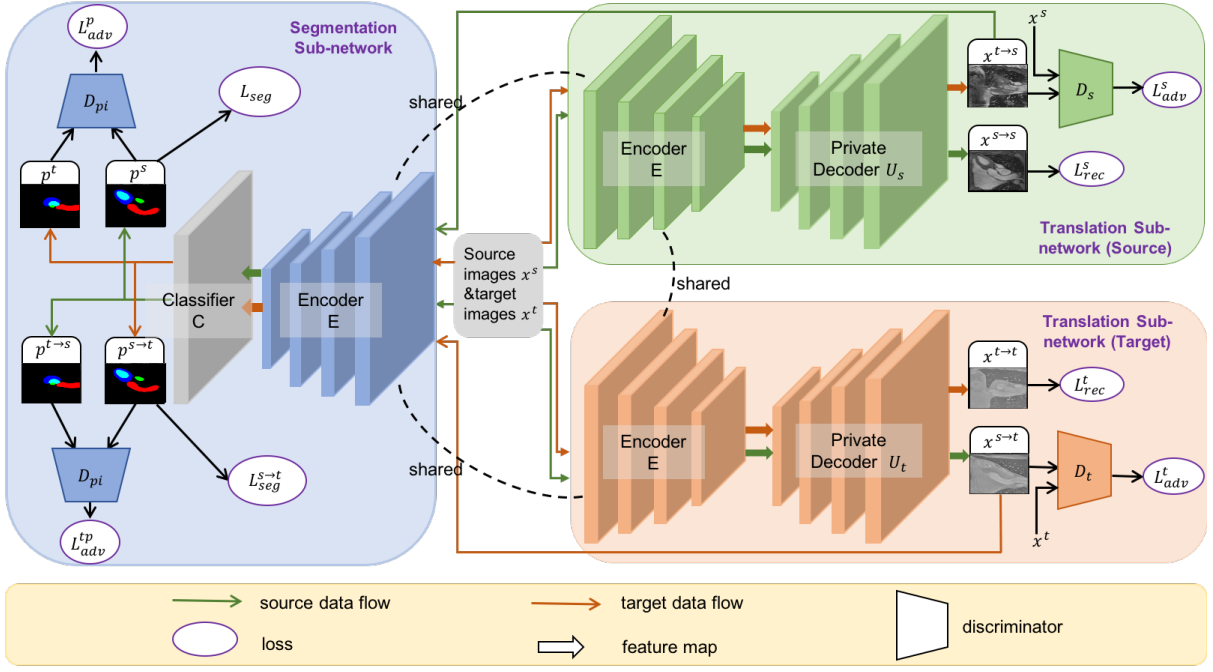}
    \caption{An overview of our proposed method. The whole network is a completely symmetric framework, which is composed of a common encoder ($E$) shared across domains, two domain-specific private decoders ($U_s$,$ U_t$) and a pixel-wise classifier ($C$). Among them, the shared encoder and a domain-specific decoder constitute a translation sub-network for reconstructing images and generating cross-domain images. In addition, the encoder and classifier form the segmentation sub-network. The green and orange blocks in the figure respectively represent the source and target domain translation sub-network, and the blue block is the segmentation sub-network.}
    \label{model}
  \end{figure*}
  
  In this section, we present technical details of our method.
  Fig.~\ref{model} illustrates the overview of our framework, which consists of a shared encoder ($E$), two domain-specific decoders ($U_s, U_t$) and a classifier ($C$). Among them, the shared encoder and a domain-specific decoder constitute a translation sub-network to reconstruct images and generate cross-domain images. The encoder and classifier form the segmentation sub-network. We focus on 1) bidirectionally aligning feature distributions, exploiting two symmetric translation sub-networks and 2) employing different styles of images to mine rich but easily underestimated semantic information. 
  In the following, we introduce our method in detail from these two aspects. The symbols used in the following sections are summarized in Table~\ref{tab:symbol}.
  
  \begin{table}[htbp]
    \centering
    \caption{SUMMARY OF SYMBOLS}
    \resizebox{0.46\textwidth}{!}{
      \begin{tabular}{ll}
      \toprule
      Symbol & Notation \\
      \midrule
      $\mathcal{X}^s, \mathcal{X}^t$ & Image sets of source and target modalities (domains)  \\
      $x^s, x^t$ & Samples of source and target modalities \\
      $x^{s \to t}, x^{t \to s}$  & Translated samples \\
      $\mathcal{Y}^s,y^s$   & Annotation set and sample label corresponding to the source domain \\
      $E$     & Shared encoder \\
      $U_s, U_t$ & Decoders for source and target domains \\
      $Z_s, Z_t$     & Embeddings of source and target domains  \\
      $D_t, D_s$ & Discriminators for source and target domains \\
      $D_{pi}$   & Discriminator for segmentation map at different levels \\
      $C_i$    & Classifier for feature map at different levels \\
      \bottomrule
      \end{tabular}}%
    \label{tab:symbol}%
  \end{table}%

  \subsection{Translation Sub-network for Feature Alignment}
  \label{subsec:generated image space}
  We aim to address the problem of domain shift in medical image segmentation, where source images ($\mathcal{X}^s$) with corresponding labels ($\mathcal{Y}^s$), and target images ($\mathcal{X}^t$) without labels, are given. Our goal is to achieve good performance in target domain by applying the model trained on source domain. However, due to the domain shift between source and target domains, the performance might significantly drop.
  Based on adversarial losses, the two translation sub-networks in our framework translate images 1) from source to target domain and 2) from target to source domain, respectively. We also introduce reconstruction loss to better maintain semantic information. Furthermore, we share the encoder among the two translation sub-networks and the segmentation sub-network to achieve the bidirectional feature alignment, which is different from previous adaptation methods. The independent decoders for source and target domains mainly pay attention to domain-specific representations, which forces the shared encoder to merely learn domain-invariant feature representations. The detailed descriptions about adversarial loss and reconstruction loss of translation sub-networks are provided in the following.
  
  \subsubsection{Adversarial Loss of Generated Space}
  Specifically, the translation sub-network is composed of the shared encoder (E) and domain-specific decoder ($U_t$). The encoder takes source images ($x^s$) as input and outputs source embedding ($Z_s$). Then the target decoder ($U_t$) takes source embedding ($Z_s$) as input and generates target-like images to fool the target discriminator ($D_t$), while the discriminator tries to distinguish translated images ($x^{s \to t}$) from target images ($x^t$). The target translation sub-network that acts as a generator, plays a minimax two-player game with the target discriminator ($D_t$). They can be optimized via the adversarial loss as below:
  \begin{equation}
    \begin{split}
    \mathcal{L}_{adv}^t(E,{U_t},{D_t}) = &{{\mathbb{E}}_{{x^t} \sim {\mathcal{X}^t}}}[\log {D_t}({x^t})] + \\
                               &{{\mathbb{E}}_{{x^s} \sim {\mathcal{X}^s}}}[\log (1 - {D_t}({U_t}(E({x^s}))))],
    \end{split}
  \end{equation}
  where the discriminator maximizes the objective in the process of optimization. Meanwhile, the encoder and decoder encourage translated images to be indistinguishable from target images, hence they make the objective continuously reduced.
  
  Besides, we can also formulate the adversarial loss of translation sub-network from target to source as below:
  \begin{equation}
    \begin{split}
    \mathcal{L}_{adv}^s(E,{U_s},{D_s}) = &{{\mathbb{E}}_{{x^s} \sim {\mathcal{X}^s}}}[\log {D_s}({x^s})] + \\
                               &{{\mathbb{E}}_{{x^t} \sim {\mathcal{X}^t}}}[\log (1 - {D_s}({U_s}(E({x^t}))))].
    \end{split}
  \end{equation}

  \subsubsection{Reconstruction Loss} To ensure translated images ($x^{s \to t}$) to preserve content information of original images and learn the style of target images, we add the target domain reconstruction loss. That is to say, we first use the encoder ($E$) and target decoder ($U_t$) to reconstruct target domain images, and then optimize them by the reconstruction loss
  \begin{equation}
  \mathcal{L}_{rec}^t(E,{U_t}) = {{\mathbb{E}}_{{x^t} \sim {\mathcal{X} ^t}}}\parallel {U_t}(E({x^t})) - {x^t}{\parallel _1}.
  \end{equation}
  
  Similarly, we also use the reconstruction loss in source translation sub-network, ensuring source decoder to learn distribution information of source domain as follows:
  \begin{equation}
    \begin{split}
    \mathcal{L}_{rec}^s(E,{U_s}) = {{\mathbb{E}}_{{x^s} \sim {\mathcal{X}^s}}}\parallel {U_s}(E({x^s})) - {x^s}{\parallel _1}.
    \end{split}
  \end{equation}
  
  So far, we obtain the optimization losses for target and source domain translation sub-networks, which can be calculated as follows:
  \begin{equation}
    \begin{split}
    \mathcal{L}_{gen}^t = \lambda _{rec}\mathcal{L}_{rec}^t + \lambda _{adv}\mathcal{L}_{adv}^t,
    \end{split}
  \end{equation}
  \begin{equation}
    \begin{split}
    \mathcal{L}_{gen}^s = \lambda _{rec}\mathcal{L}_{rec}^s + \lambda _{adv}\mathcal{L}_{adv}^s,
    \end{split}
  \end{equation}
  where $\lambda _{rec}$, $\lambda _{adv}$ are hyper-parameters, which are used to adjust the weight of each component. The corresponding values in the experiment are 1.0, 0.01, respectively.


  \subsection{Segmentation Sub-network for Semantic Mining}
  \label{subsec:semantic space}
  
  The previous methods~\cite{chen2020unsupervised,chen2019synergistic} utilize translated source images to learn the semantic information. Since translated source images might miss some semantic information in the image translation stage, caused by the instability of GAN-based methods~\cite{goodfellow2014generative, isola2017image}, we utilize both translated source images and original source images to train the segmentation sub-network. Furthermore, the raw target images and translated target images are employed to conduct the adversarial task, which can help the segmentation sub-network focus more on the common semantic knowledge of source and target domains. This sub-network includes two kinds of losses, \ie, segmentation loss and adversarial loss.
  
  \subsubsection{Segmentation Loss}
   Since all images from source domain have corresponding labels, we feed source images ($x^s$) into the encoder ($E$) and pixel-level classifier (C) to generate prediction maps. Then we optimize the whole segmentation sub-network by the segmentation loss. Inspired from the SIFA~\cite{chen2020unsupervised}, we also apply the idea of deep supervision. We add an additional classifier for prediction of lower feature maps. To overcome the class imbalance issue between relative small-sized foreground and large-sized background, we employ a sum of soft Dice and weighted cross-entropy (CE) loss to train the segmentation sub-network as follows:
  \begin{equation}
    \begin{split}
    \mathcal{L}_{seg}^s(E,{C_i}) = &\mathcal{L}_\text{CE}({C_i}(E({x^s})),{y^s}) + \\
          &\mathcal{L}_\text{Dice}({C_i}(E({x^s})),{y^s}), i=1,2.
    \end{split}
  \end{equation}

  In addition, the translated images ($x^{s \to t}$) maintain some content information of original images, and they can inherit the ground truth of source images, thus these images can be used for supervised training of the segmentation sub-network. We feed the translated images ($x^{s \to t}$) into segmentation sub-network and calculate the segmentation loss with source domain labels. The cross-domain segmentation loss, which is similar to source domain segmentation loss, is calculated as follows:
  \begin{equation}
    \begin{split}
    \mathcal{L}_{seg}^{s \to t}(E,{C_i}) = & \mathcal{L}_\text{CE}({C_i}(E({x^{s \to t}})),{y^s}) + \\
                             & \mathcal{L}_\text{Dice}({C_i}(E({x^{s \to t}})),{y^s}), i=1,2.
    \end{split}
  \end{equation}
  
  \subsubsection{Adversarial Loss of Semantic Space}
  It is known that segmentation outputs of source and target domains share a significant amount of similarity, \eg, spatial layout and local context. Also, source domain can get corresponding labels, while there is no label information in target domain. Nevertheless,  prediction maps can be utilized to aid the segmentation sub-network to mine the shared semantic knowledge between source and target domains.
  Specifically, we input target images into the segmentation sub-network, and the prediction maps, together with the source prediction maps, are feed into the discriminators ($D_{pi}$), which aims to distinguish the prediction maps from source domain or target domain. Differently, the segmentation sub-network fools the discriminators ($D_{pi}$). Therefore, the adversarial loss in the semantic space of real images is defined as:
  \begin{equation}
    \begin{split}
    \mathcal{L}_{adv}^p(E,{C_i},{D_{pi}}) = &{{\mathbb{E}}_{{x^s} \sim {\mathcal{X}^s}}}[\log {D_{pi}}({C_i}(E({x^s})))] + \\
                                  &{{\mathbb{E}}_{{x^t} \sim {\mathcal{X}^t}}}[\log (1 - {D_{pi}}({C_i}(E({x^t}))))].
    \end{split}
  \end{equation}
  
  Similarly, translated source images and translated target images are available from the translation sub-networks, and translated source images inherit source labels. We add the corresponding adversarial loss in the semantic space of generated images, which can be described as:
  \begin{equation}
    \begin{split}
    \mathcal{L}_{adv}^{tp}(E,{C_i},{D_{pi}}) = &{\mathbb{E}}[\log {D_{pi}}({C_i}(E({x^{s \to t}})))] + \\
                                     &{\mathbb{E}}[\log (1 - {D_{pi}}({C_i}(E({x^{t \to s}}))))].
    \end{split}
  \end{equation}
  
  To sum up, in order to mine more semantic information, the whole adversarial loss in the semantic space can be expressed as:
  \begin{equation}
    \begin{split}
    \mathcal{L}_{adv}^{sec}(E,{C_i},{D_{pi}}) = \mathcal{L}_{adv}^p(E,{C_i},{D_{pi}}) + \mathcal{L}_{adv}^{tp}(E,{C_i},{D_{pi}}).
    \end{split}
  \end{equation}

  \subsection{The Overall Loss}
  
  In summary, the whole objective of our network for unsupervised domain adaptation is formulated as follows:
  \begin{equation}
    \begin{split}
    \mathcal{L} = &\lambda _{gen}(\mathcal{L}_{gen}^t + \mathcal{L}_{gen}^s) +\\ &\lambda _{seg}(\mathcal{L}_{seg}^s + \mathcal{L}_{seg}^{s \to t}) + \lambda _{adv}^{sec}\mathcal{L}_{adv}^{sec},
    \end{split}
  \end{equation}
  where $\lambda _{gen}$, $\lambda _{seg}$, and $\lambda _{adv}^{sec}$ are hyper-parameters adjusting the weight of each module. The corresponding values in our experiment are set to 1.0, 1.0 and 0.1, respectively.

  \subsection{Network Architecture}
  \label{subsec:Network architecture}
  The whole framework consists of the segmentation sub-network and two translation sub-networks. Among them, these three parts share an encoder. The segmentation sub-network, which is built on Deeplab-ResNet50, includes the encoder (E) and classifiers ($C_i$). Two domain-specific decoders have the same architecture, but they do not share any weights. They are composed of three residual blocks and three deconvolution layers. There are three kinds of discriminators, which are used to distinguish true and generated images of source domain, true and generated images of target domain, and different domains of output prediction maps. All of them have the same architecture, including four convolution layers, but do not share any parameters with each other.

  \section{Experiments}
  \label{sec:experiments}
  In this section, we evaluate the effectiveness of our proposed method from both qualitative and quantitative perspectives. Specifically, we compare our method with several state-of-the-art methods and perform some ablation experiments to investigate the effect of various constraints on model performance.
  
  \subsection{Datasets and Implementation Details}
  \label{subsec:dataset}
  \subsubsection{Datasets}
  In our experiment, we utilize two datasets (\ie, Cardiac dataset~\cite{zhuang2016multi} and BraTS dataset~\cite{menze2014multimodal}) to validate the efficacy of our method.
  The first dataset is the Multi-Modality Whole Heart Segmentation (MMWHS) Challenge 2017 dataset~\cite{zhuang2016multi}, which consists of unpaired 20 MRI and 20 CT volumes from different clinical sites. We complete adaptation experiments both in the ``MRI to CT'' direction and in the ``CT to MRI'' direction. The label mask of cardiac images is manually delineated by the professional radiologists. The segmentation labels contain four cardiac structures, which are the left ventricle myocardium (LVM), left atrium blood cavity (LAB), left ventricle blood cavity (LVB) and ascending aorta (AA).
  The other dataset is the Multi-Modality Brain Tumor Segmentation Challenge 2018 dataset~\cite{menze2014multimodal}, which is composed of four modalities of MRI imaging: FLAIR, T1, T1CE, and T2. In the experiment, T2 is treated as the source domain and the other modalities are considered as the target domains. We aim to segment the whole brain tumor of target domains in an unpaired way. For both datasets, we randomly select 80\% of patient data as the training set and 20\% as the test set. The image intensity is first normalized by subtracting the mean intensity and dividing by the standard deviation, and then transferred to range [-1, 1]. We perform the random crop, rotation and other augmentation operations in the data pre-processing stage.
  
  \subsubsection{Implementation Details}
  We implement our model with PyTorch on a Tesla V100 with 32 GB memory. The model on the Cardiac dataset is trained for $2 \times {10^4}$ iterations and the batch size is set to 8. The full training process takes 10 hours. On the BraTS dataset, we train the model for $6 \times {10^4}$ iterations with a batch size of 8, and the full training process takes 24 hours. For both tasks, the shared encoder ($E$) and classifiers ($C_i$) are trained with the Adam optimizer with learning rate of $2 \times {10^{-4}}$. The two domain-specific decoders are trained with the Adam optimizer using the learning rate of $1 \times {10^{-3}}$. Differently, the domain-specific discriminators ($D_s$ and $D_t$) are trained with the Adam optimizer with learning rate of $1 \times {10^{-4}}$, and the discriminators of semantic space ($D_{pi}$) are trained with the Adam optimizer with learning rate of $5 \times {10^{-5}}$.
  
  \subsection{Evaluation Metrics}
  \label{subsec:evaluation metrics}
  There are three evaluation metrics in our experiment, including the Dice similarity coefficient (Dice), the average surface distance (ASD) and Hausdorff distance (HD). Dice is mainly used to calculate the similarity between prediction map and ground truth. Higher Dice value indicates better segmentation performance. ASD is used to calculate the average distances between the surface of the prediction mask and ground truth in 3D. And HD is introduced to measure the extreme distance between two sets of points. Lower ASD and HD values indicate better performance. Being consistent with previous works~\cite{chen2020unsupervised,zouunsupervised}, we choose Dice and ASD as evaluation metrics on the Cardiac dataset, and choose Dice and HD as evaluation metrics on the BraTS dataset.
  \begin{table*}[!t]
  \centering
  \caption{Comparison of different methods on the Cardiac dataset. The \textbf{bold} number highlights the best performance.}
  \label{cardiac performance comparison}
  \resizebox{0.9\textwidth}{!}{
  \scriptsize
  \renewcommand{\arraystretch}{0.9}
  \begin{tabular}{c|ccccc|ccccc}
  \toprule
  \multicolumn{11}{c}{Cardiac MRI $\rightarrow$ Cardiac CT} \\
  \midrule
  \multirow{2}[4]{*}{Method} & \multicolumn{5}{c|}{Dice (\%)}         & \multicolumn{5}{c}{ASD (voxel)} \\
  \cmidrule{2-11}          & AA    & LAC   & LVC   & MYO   & Average & AA    & LAC   & LVC   & MYO   & Average \\
  \midrule
  Supervised training  & 76.14  & 89.82  & 91.37  & 85.47  & 85.70  & 3.39  & 3.64  & 2.44  & 2.19  & 2.91  \\
  No adaptation & 37.02  & 33.71  & 0.27  & 4.65  & 18.92  & 28.64  & 8.56  & 58.50  & 28.29  & 31.00  \\
  \midrule
  PnP-AdaNet~\cite{dou2019pnp} & 74.00  & 68.90  & 61.90  & 50.80  & 63.90  & 12.80  & 6.30  & 17.40  & 14.70  & 12.80  \\
  AdaOutput~\cite{tsai2018learning} & 65.20  & 76.60  & 54.40  & 43.60  & 59.90  & 17.90  & \textbf{5.50}  & 5.90  & 8.90  & 9.60  \\
  CycleGAN~\cite{zhu2017unpaired} & 73.80  & 75.70  & 52.30  & 28.70  & 57.60  & 11.50  & 13.60  & 9.20  & 8.80  & 10.80  \\
  CyCADA~\cite{hoffman2018cycada} & 72.90  & 77.00  & 62.40  & 45.30  & 64.40  & 9.60  & 8.00  & 9.60  & 10.50  & 9.40  \\
  SIFA~\cite{chen2020unsupervised}  &81.30  & 79.50  & 73.80  & 61.60  & 74.10  & 7.90  & 6.20  & 5.50  & 8.50 & 7.00  \\
  DSFN~\cite{zouunsupervised}  & \textbf{84.70} & 76.90  & 79.10  & 62.40  & 75.80  & N/A     & N/A    & N/A    & N/A     & N/A \\
  Ours  & 79.92 & \textbf{84.76} & \textbf{82.77} & \textbf{66.52} & \textbf{78.50} & \textbf{7.68}  & 6.65 & \textbf{3.77} & \textbf{5.59}  & \textbf{5.92} \\
  \bottomrule
  \end{tabular}}%
  \vspace{-0.2cm}
  \end{table*}%
          
  \begin{table*}[!h]
  \centering
  \resizebox{0.9\textwidth}{!}{
  \scriptsize
  \renewcommand{\arraystretch}{0.9}
  \begin{tabular}{c|ccccc|ccccc}
  \toprule
  \multicolumn{11}{c}{Cardiac CT $\rightarrow$ Cardiac MRI} \\
  \midrule
  \multirow{2}[4]{*}{Method} & \multicolumn{5}{c|}{Dice (\%)}         & \multicolumn{5}{c}{ASD (voxel)} \\
  \cmidrule{2-11}          
  & AA    & LAC   & LVC   & MYO   & Average & AA    & LAC   & LVC   & MYO   & Average \\
  \midrule
  Supervised training  & 82.24  & 86.33  & 91.51  & 78.28  & 84.59  & 3.44  & 1.82  & 1.52  & 1.72  & 2.13  \\
  No adaptation & 0.08  & 9.31  & 12.50  & 0.73  & 5.66  & 57.92  & 18.09  & 16.12  & 16.81  & 27.23  \\
  \midrule
  PnP-AdaNet~\cite{dou2019pnp} & 43.70  & 47.00  & 77.70  & 48.60  & 54.30  & 11.40  & 14.50  & 4.50  & 5.30  & 8.90  \\
  AdaOutput~\cite{tsai2018learning} & 60.80  & 39.80  & 71.50  & 35.50  & 51.90  & 5.70  & 8.00  & 4.60  & 4.60  & 5.70 \\
  CycleGAN~\cite{zhu2017unpaired} & 64.30  & 30.70  & 65.00  & 43.00  & 50.70  & 5.80  & 9.80  & 6.00  & 5.00  & 6.60  \\
  CyCADA~\cite{hoffman2018cycada} & 60.50  & 44.00  & 77.60  & 47.90  & 57.50  & 7.70  & 13.90  & 4.80  & 5.20  & 7.90  \\
  SIFA~\cite{chen2020unsupervised}  & 65.30  & 62.30  & \textbf{78.90}  & 47.30  & 63.40  & 7.30  & 7.40  & \textbf{3.80}  & 4.40  & 5.70 \\
  Ours  & \textbf{71.29} & \textbf{66.23}  & 76.20 & \textbf{52.07} & \textbf{66.45} & \textbf{4.44} & \textbf{7.30}  & 5.46  & \textbf{4.25} & \textbf{5.36} \\
  \bottomrule
  \end{tabular}}%
  \vspace{-0.2cm}
  \end{table*}%

  \begin{figure*}[!h]
  \centering
  \includegraphics[width=0.9\textwidth]{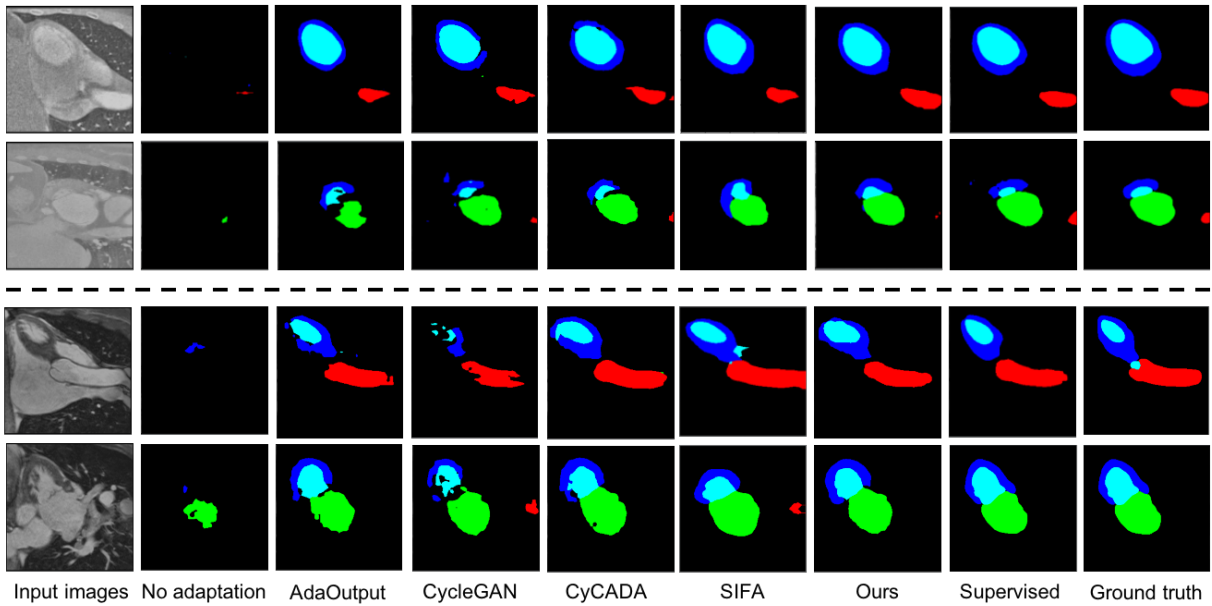}
  \caption{Cardiac segmentation results of different unsupervised domain adaptation methods. The first two rows are CT images in the ``MRI to CT'' task, and the last two rows are MR images in the reverse task.}
  \label{cardiac visual results}
  \end{figure*}
  \subsection{Comparison with the State-of-the-art Methods}
  \label{subsec:performance comparison}
  \begin{table*}[!t]
  \centering
  \caption{Comparison of different methods on the BraTS dataset. The \textbf{bold} number highlights the best performance.}
  \label{brats performance comparison}%
  \resizebox{0.9\textwidth}{!}{
  \tiny
  \renewcommand{\arraystretch}{0.9}
  \begin{tabular}{c|cccc|cccc}
  \toprule
  \multirow{2}[4]{*}{Method} & \multicolumn{4}{c|}{Dice (\%)} & \multicolumn{4}{c}{Hausdorff Distance (mm)} \\
  \cmidrule{2-9}          & T1    & FLAIR & T1CE  & Average & T1    & FLAIR & T1CE  & Average \\
  \midrule
  Supervised training  & 73.17  & 85.58  & 72.56  & 77.10  & 9.47  & 4.57  & 9.18  & 7.74  \\
  No adaptation & 4.17  & 65.16  & 6.28  & 27.70  & 55.67  & 28.00  & 49.77  & 39.56  \\
  \midrule
  AdaOutput~\cite{tsai2018learning} & 42.60  & 67.80  & 33.10  & 47.80  & 23.40  & 16.60  & 28.80  & 22.90  \\
  CycleGAN~\cite{zhu2017unpaired} & 38.10  & 63.30  & 42.10  & 47.80  & 25.40  & 17.20  & 23.20  & 21.90  \\
  CyCADA~\cite{hoffman2018cycada} & 49.60  & 72.00  & 51.70  & 57.80  & 20.20  & 14.90  & 18.40  & 17.80  \\
  SIFA~\cite{chen2020unsupervised}  & 51.70  & 68.00  & 58.20  & 59.30  & 19.60  & 16.90  & 15.01  & 17.10  \\
  DSFN~\cite{zouunsupervised}  & 57.30  & 78.90  & \textbf{62.20}  & 66.10  & 17.50  & 13.80  & 15.50  & 15.60  \\
  Ours  & \textbf{57.70} & \textbf{81.79} & 62.04  & \textbf{67.18} & \textbf{14.24} & \textbf{8.62} & \textbf{13.70} & \textbf{12.19} \\
  \bottomrule
  \end{tabular}}%
  \vspace{-0.2cm}
  \end{table*}%
              
  \begin{figure*}[!h]
  \centering
  \includegraphics[width=0.96\textwidth]{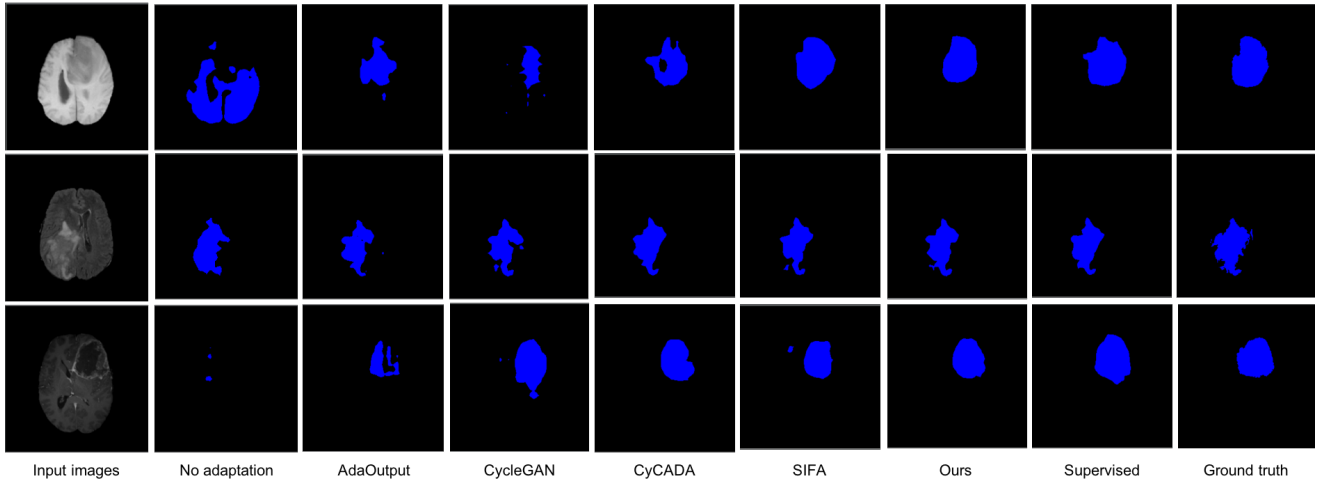}
  \caption{Brain tumor segmentation results of different methods in the unsupervised domain adaptation task. The input images from top to bottom are from T1, FLAIR, and T1CE modality, respectively.}
  \label{brats visual results}
  \end{figure*}
  We now report the quantitative results of our method and the state-of-the-art (SOTA) methods in the Cardiac and BraTS segmentation tasks.
  
  In Table~\ref{cardiac performance comparison}, we visualize the adaptation results in the task ``MRI to CT'' and ``CT to MRI'' of the Cardiac dataset.
  First, the result of ``supervised training'' can be regarded as the upper bound of unsupervised domain adaption methods. As seen in Table~\ref{cardiac performance comparison}, it achieves Dice of 85.70\% in the ``MRI to CT'' task and 84.59\% in the reverse task.
  ``No adaptation'' denotes the evaluation results on CT images using the model trained on MR images or vice versa, which can be taken as the lower bound of unsupervised domain adaption methods. 
  When compared to ``supervised training'', we can observe a drastic drop in performance due to the distribution discrepancy. For example, the Dice is reduced by 66.78\% in the ``MRI to CT'' task and 78.93\% in the ``CT to MRI'' task.
  
  We choose several recently proposed SOTA methods for comparison. In particular, PnP-AdaNet~\cite{dou2019pnp} aligns the feature spaces of source and target domains in an unsupervised manner;
  The AdaOutput~\cite{tsai2018learning} conducts the adversarial task between source and target prediction maps;
  The CycleGAN~\cite{zhu2017unpaired} performs data augmentation in the target domain via image-to-image translation methods;
  The CyCADA~\cite{hoffman2018cycada} adapts representations at both the image-level and feature-level;
  The SIFA~\cite{chen2020unsupervised} conducts synergistic alignment of domain from both image and feature perspectives;
  And DSFN~\cite{zouunsupervised} introduces a dual-scheme fusion network to reduce the domain gap.
  Among these methods, PnP-AdaNet and CycleGAN are the feature-alignment-based method and image-translation-based method, respectively. CyCADA, SIFA and DSFN are the joint learning methods of feature alignment and image translation. In particular, our method also belongs to the joint learning method.
  Different from other joint learning methods, from the feature alignment perspective, we propose the bidirectional feature alignment using two symmetric translation sub-networks. From the image translation perspective, we exploit all styled images to train the segmentation sub-network via segmentation loss and adversarial loss.
  
  According to Table~\ref{cardiac performance comparison}, we can observe that 1) joint learning paradigm outperforms the independent image-translation or feature-alignment manner. The main reason is that image translation can help achieve better feature alignment in joint learning paradigm and vice versa. For example,
  PnP-AdaNet and CycleGAN perform worser than joint learning methods of feature alignment and image translation (\ie, CyCADA, SIFA, and DSFN). As for AdaOutput, it adopts the alignment in the semantic space, which also belongs to the category of independent methods. Its test result is also in accordance with this observation.
  2) Compared to the existing joint learning methods, our proposed method shows better performance, which is attributed to the proposed bidirectional alignment and semantic mining schemes. As seen in Table~\ref{cardiac performance comparison},
  compared to CyCADA, our method significantly outperforms CyCADA by 14.10\% in ``MRI to CT'' and 8.95\% in ``CT to MRI'', which demonstrates the effectiveness of the shared encoder between segmentation sub-network and translation sub-networks.
  In addition, our method improves the Dice by 4.40\% (``MRI to CT'') and 3.05\% (``CT to MRI'') when compared to SIFA. The main reason might be that we fully exploit all styled images to incorporate more semantic knowledge.
  As for DSFN, our method outperforms DSFN by 2.70\% in ``MRI to CT'' since we implement bidirectional feature alignment via two symmetric translation sub-networks.
  Some typical segmentation results of these adaptation methods on the Cardiac dataset are shown in Fig.~\ref{cardiac visual results}.
  
  \begin{table*}[!t]
    \centering
    \caption{Experimental results of ablation study. The \textbf{bold} number highlights the best performance.}
    \label{tab:ablation study}
    \resizebox{0.95\textwidth}{!}{
      \scriptsize
      \renewcommand{\arraystretch}{0.9}   
      \begin{tabular}{l|ccccc|ccccc}
      \toprule
      \multicolumn{11}{c}{Cardiac MRI $\rightarrow$ Cardiac CT} \\
      \midrule
      \multirow{2}[4]{*}{Method} & \multicolumn{5}{c|}{Dice (\%)}         & \multicolumn{5}{c}{ASD (voxel)} \\
  \cmidrule{2-11}          & AA    & LAC   & LVC   & Myo   & Average & AA    & LAC   & LVC   & Myo   & Average \\
      \midrule
      $\mathcal{L}_{seg}^s$     & 37.02  & 33.71  & 0.27  & 4.65  & 18.92  & 28.64  & 8.56  & 58.50  & 28.29  & 31.00  \\
      $\mathcal{L}_{gen}^t+\mathcal{L}_{seg}^{s \to t}$     & 71.11  & 71.21  & 56.37  & 46.67  & 61.34  & \textbf{6.11}  & 6.84  & 9.77  & 8.93  & 7.91  \\
      $\mathcal{L}_{gen}^t+\mathcal{L}_{seg}^{s \to t}+\mathcal{L}_{gen}^s$     & \textbf{82.11}  & 76.53  & 70.36  & 42.75  & 67.94   & 7.48  & 7.27   & 6.95   & 7.68  & 7.34   \\
      $\mathcal{L}_{gen}^t+\mathcal{L}_{seg}^{s \to t}+\mathcal{L}_{gen}^s+\mathcal{L}_{seg}^s$     & 81.04 & 77.03  & 69.42  & 51.59  & 69.77  & 7.82  & \textbf{4.98} & 10.09  & 6.24  & 7.28  \\
      $\mathcal{L}_{gen}^t+\mathcal{L}_{seg}^{s \to t}+\mathcal{L}_{gen}^s+\mathcal{L}_{seg}^s+\mathcal{L}_{adv}^{sec}(all)$     & 79.92  & \textbf{84.76} & \textbf{82.77} & \textbf{66.52} & \textbf{78.50} & 7.68  & 6.65  & \textbf{3.77} & \textbf{5.59}  & \textbf{5.92} \\
      \bottomrule
      \end{tabular}}%
      \vspace{-0.2cm}
  \end{table*}%

  \begin{table*}[htbp]
    \centering
    \resizebox{0.95\textwidth}{!}{
      \scriptsize
      \renewcommand{\arraystretch}{0.9}   
      \begin{tabular}{l|ccccc|ccccc}
      \toprule
      \multicolumn{11}{c}{Cardiac CT $\rightarrow$ Cardiac MRI} \\
      \midrule
      \multirow{2}[4]{*}{Method} & \multicolumn{5}{c|}{Dice (\%)}         & \multicolumn{5}{c}{ASD (voxel)} \\
  \cmidrule{2-11}          & AA    & LAC   & LVC   & Myo   & Average & AA    & LAC   & LVC   & Myo   & Average \\
      \midrule
      $\mathcal{L}_{seg}^s$     & 0.08  & 9.31  & 12.50  & 0.73  & 5.66  & 57.92  & 18.09  & 16.12  & 16.81  & 27.23  \\
      $\mathcal{L}_{gen}^t+\mathcal{L}_{seg}^{s \to t}$     & 57.19  & 22.43  & 67.92  & 35.77  & 45.83  & 6.53  & 13.38  & 5.35  & 5.23  & 7.62  \\
      $\mathcal{L}_{gen}^t+\mathcal{L}_{seg}^{s \to t}+\mathcal{L}_{gen}^s$     & 56.50    & 47.61    & 75.12   & 35.30    & 53.63   & 6.17   & 10.73   & 4.89   & 5.25    & 6.76    \\
      $\mathcal{L}_{gen}^t+\mathcal{L}_{seg}^{s \to t}+\mathcal{L}_{gen}^s+\mathcal{L}_{seg}^s$     & 57.70  & 50.86  & 75.79  & 32.54  & 54.22  & 7.80  & 8.68  & \textbf{4.64}  & 4.68  & 6.45  \\
      $\mathcal{L}_{gen}^t+\mathcal{L}_{seg}^{s \to t}+\mathcal{L}_{gen}^s+\mathcal{L}_{seg}^s+\mathcal{L}_{adv}^{sec}(all)$     & \textbf{71.29} & \textbf{66.23}  & \textbf{76.20} & \textbf{52.07} & \textbf{66.45} & \textbf{4.44} & \textbf{7.30}  & 5.46  & \textbf{4.25} & \textbf{5.36} \\
      \bottomrule
      \end{tabular}}%
      \vspace{-0.2cm}
  \end{table*}%
  
  \begin{figure*}
    \centering
    \subfigure[MRI to CT]{
    \includegraphics[width=0.48\textwidth, height=0.25\textwidth]{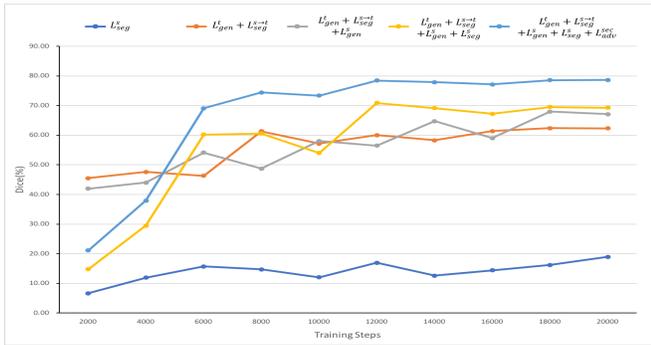}
    }
    \subfigure[CT to MRI]{
    \includegraphics[width=0.48\textwidth, height=0.25\textwidth]{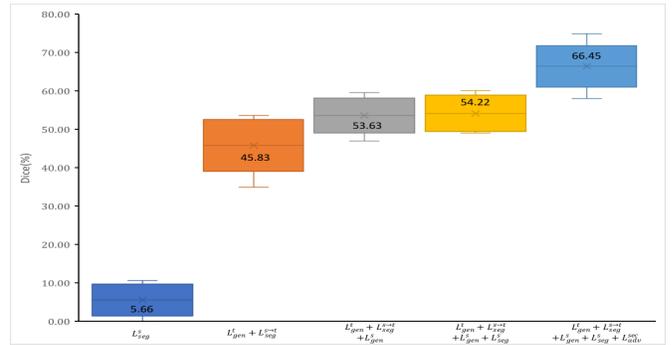}
    }
    \caption{(a) The intermediate segmentation performance of the proposed network and its variant models after different training steps in ``MRI to CT''. (b) The segmentation result boxplot with different ablation experimental settings in ``CT to MRI''. The confidence intervals are generated based on the different volumes in the test sets.}
    \label{ablation}
  \end{figure*}
  
  In Table~\ref{brats performance comparison}, we display the segmentation results on the BraTS dataset. Similar with that in the Cardiac dataset, we first report the experimental results of ``supervised training'' and ``no adaptation'' methods. Due to the domain gap, the Dice is reduced by 49.40\% and HD is increased by 31.82.
  Since the appearance of FLAIR and T2 modality is similar, we achieve the Dice of 65.16\% and the HD of 28.00, respectively, without adaptation. Differently, T1 and T1CE have a significant distribution discrepancy with T2 modality, the Dice are only 4.17\% and 6.28\%, respectively.
  From Table~\ref{brats performance comparison}, we can observe that the performance of all adaptation methods on this dataset shows the same trend as the Cardiac dataset.
  Some typical segmentation results of these adaptation methods for the BraTS dataset are illustrated in Fig.~\ref{brats visual results}. As observed, our segmentation results are closer to supervised results and better than other methods.
  
  \subsection{Ablation Study}
  \label{subsec:ablation study}
  We perform ablation experiments on Cardiac dataset to verify the impact of various constraints on the entire network. The experimental results are displayed in Table~\ref{tab:ablation study}. Overall, we focus on bidirectionally aligning feature distributions between source and target domains from the generated image space and learning common semantic knowledge in the semantic space.
  
  \subsubsection{Effectiveness of unidirectional alignment}
  We perform an experiment that only exploits unidirectional alignment via the ``from source to target'' translation sub-network and apply the translated segmentation loss ($\mathcal{L}_{gen}^t+\mathcal{L}_{seg}^{s \to t}$ ). Compared to the result of ``no adaptation'', the average Dice increases by 42.42\% and the average ASD is reduced by 23.09 in the ``MRI to CT'' task. Besides, in the ``CT to MRI'' task, the average Dice increases by 40.17\% and the average ASD is reduced by 19.61.
  
  \subsubsection{Effectiveness of bidirectional alignment}
  We perform an experiment to conduct bidirectional feature alignment ($\mathcal{L}_{gen}^t +\mathcal{L}_{seg}^{s \to t}+\mathcal{L}_{gen}^s$) by adding a symmetric ``from target to source'' translation sub-network. The two translation sub-networks consist of the shared encoder and private decoders, where two private decoders learn domain-specific representations by reconstruction losses and the shared encoder focuses on learning domain-invariant representations.
  In this way, our average Dice increases by 6.60\% in ``MRI to CT'' and 7.80\% in ``CT to MRI'', compared to ``$\mathcal{L}_{gen}^t+\mathcal{L}_{seg}^{s \to t}$'' (unidirectional alignment).

  \subsubsection{Effectiveness of source segmentation}
  Due to the instability of GAN-based methods~\cite{goodfellow2014generative, isola2017image}, translated source images might miss some semantic information of source images. As a result, we feed source images into the segmentation sub-network to make up for lost semantic information ($\mathcal{L}_{gen}^t+\mathcal{L}_{seg}^{s \to t}+\mathcal{L}_{gen}^s+\mathcal{L}_{seg}^s$). The Dice increases by 1.83\% in ``MRI to CT'' and 0.59\% in ``CT to MRI'' when compared to ``$\mathcal{L}_{gen}^t +\mathcal{L}_{seg}^{s \to t}+\mathcal{L}_{gen}^s$''.
  
  \subsubsection{Effectiveness of adversarial loss in semantic space}
  It is known that not only original source images are accessible to corresponding labels, but translated source images can inherit source labels. Being aware this fact, we use original images and generated images to respectively perform adversarial learning ($\mathcal{L}_{gen}^t+\mathcal{L}_{seg}^{s \to t}+\mathcal{L}_{gen}^s+\mathcal{L}_{seg}^s+\mathcal{L}_{adv}^{sec}$) in the semantic space, which can exploit all styled images to explore more common semantic information.
  By this way, the average Dice increases to 78.50\% (``MRI to CT'') and 66.45\% (``CT to MRI''), and ASD is reduced to 5.92 (``MRI to CT'') and 5.36 (``CT to MRI'').
  Moreover, Fig.~\ref{ablation} shows the intermediate segmentation performance of the proposed network and its variant models after different training steps in the ``MRI to CT'' task and shows the ``CT to MRI'' segmentation result boxplot from different volumes in the MR test set. As observed, each module proposed in our method is useful for performance improvement.
  
  
  
  \begin{figure}[!t]
    \centering
    \includegraphics[width=0.48\textwidth]{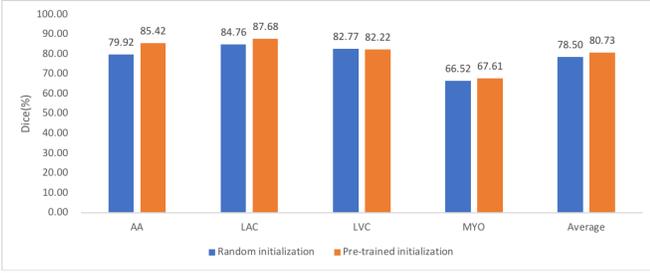}
    \caption{Comparison of the random initialization (blue) and the pre-trained initialization (orange).}
    \label{initial}
    \vspace{-0.2mm}
  \end{figure}
  
  \begin{table}[!t]
    \centering
    \caption{Comparison of different parameter setting and our method. The \textbf{bold} number highlights the best performance.}
      \footnotesize
      \renewcommand{\arraystretch}{0.9}
      \begin{tabular}{c|c|c}
        \toprule
        \multicolumn{3}{c}{Cardiac MRI $\rightarrow$ Cardiac CT} \\
        Method & Dice (\%) & ASD (voxel) \\
        \midrule
        Shared Decoders & 44.93  & 10.50  \\
        Unshared Encoders & 73.90      & 6.90      \\
        Unshared Discriminators & 76.14  & 6.04  \\
        Ours  & \textbf{78.50} & \textbf{5.92} \\
        \midrule
        \multicolumn{3}{l}{``Ours'' : private decoders, shared encoders, and shared discriminators} \\ 
        \bottomrule
       \end{tabular}%
    \label{tab:discussion}
  \end{table}
  
  \section{Discussion}
  \label{sec:discussion}
  In this part, we mainly discuss different parameter settings in each module and our method. We report the experimental results in Fig.~\ref{initial} and Table~\ref{tab:discussion}.
  
  \subsubsection{Initialization of network parameters}
  In the experiment, there are two ways to initialize the network parameters, \ie, 1) random initialization and 2) initialization with a pre-trained network.
  Specifically, for the latter, we use the pre-trained model ResNet50 on the ImageNet dataset~\cite{deng2009imagenet}. In the experiment, we only reuse the parameters of layer1, layer2, layer3, layer4 in the pre-trained ResNet50 model, since our segmentation network replaces the last fully connected layer with a convolutional layer and changes the input channel to one.
  As verified by experiment, due to the initialization using pre-trained model, the encoder can extract more representative features and the average Dice is improved to 80.73\% compared with random initialization, as shown in Fig.~\ref{initial}.
  However, in order to make a fair comparison with other methods, we adopt the results of random initialization in this paper.

  \subsubsection{Evaluation of domain-specific decoders}
  In our proposed network, there are two translation sub-networks, which are composed of a common shared encoder and two private decoders. The private decoders are used to learn domain-specific representations via reconstruction losses and the shared encoder is used to learn domain-invariant representations via adversarial losses.
  If we share the decoders of translation sub-networks in our framework, it means that source and target domains apply a common translation network. The shared decoders could not learn domain-specific representations and could not encourage the shared encoder to learn the domain-invariant representations.
  As verified by experiment (Shared Decoders) in Table~\ref{tab:discussion}, the result of shared decoders is poorer than private decoders in ``ours''. It only achieves the Dice of 44.93\% and the ASD of 10.50.
  
  \subsubsection{Evaluation of shared encoders}
  We share the encoder between the segmentation sub-network and two translation sub-networks. In this way, we achieve the bidirectional feature alignment.
  If the segmentation sub-network does not share the encoder with two translation sub-networks, it means that they are independent with each other. The network could not achieve the feature alignment from the generated image space.
  As observed in Table~\ref{tab:discussion}, the Dice is reduced by 4.60\% in ``unshared encoders'' when compared to shared encoders in ``ours''. 
  
  \subsubsection{Effectiveness of the shared discriminators}
  In our experiments, we use two sets of discriminators for adversarial learning in the semantic space. We share the parameters of these two sets of discriminators.
  Between them, one set of discriminators is used to distinguish prediction maps of real images which domain they come from. In this set, since the source domain is annotated, its label is a positive category and the target domain is a negative category.
  The other set of discriminators are used to distinguish which domain prediction maps of generated images come from. Since translated source images have corresponding labels, they are marked as a positive class, and translated target images are marked as a negative class.
  If the parameters are shared between the two sets of discriminators, it means the real and generated images use the same classifier. In addition, source images and translated source images are grouped into the same category, and target images and translated target images are grouped into the other category, which exactly meets the requirements of semantic consistency.
  As seen in Table~\ref{tab:discussion}, sharing parameters of discriminators in ``ours'' improves the Dice by 2.36\% when compared to not sharing parameters (\ie, Unshared Discriminators).
  
  \section{Conclusion}
  \label{sec:conclusion}
  In this paper, we proposed a novel symmetric network of unsupervised domain adaptation for medical image segmentation. Our model not only achieves the bidirectional feature alignment via two symmetric translation sub-networks, but also sufficiently explores the semantic information from different style images. Through the experiments, our network shows significant advantages compared to the state-of-the-art methods on both the Cardiac and BraTS datasets.
  
  Besides, our method can be further improved to better generalize in the target domain. For example, if we can get pseudo-labels of target domain or combine the self-training paradigm, more semantic information from target domain could be explored to achieve better segmentation performance. We will explore this direction in our future work.
%
%
%

\ifCLASSOPTIONcaptionsoff
  \newpage
\fi



%

\bibliographystyle{IEEEtran}
\bibliography{egpaper_final}

\balance

\end{document}